\documentclass[onecollarge,runningheads]{svjour2}
\smartqed  
\usepackage{graphicx}
\usepackage{amsmath,amssymb,bm}
\usepackage{latexsym}
\begin{document}

\title{Theory of defect dynamics in graphene: defect groupings and their stability\thanks{Work financed by the Spanish Ministry of Science and Innovation under grants 
FIS2008-04921-C02-01 and FIS2008-04921-C02-01.} }
\author{L.L. Bonilla \and A. Carpio }
\institute{L.L. Bonilla \at G. Mill\'an Institute, Fluid Dynamics, Nanoscience and Industrial
Mathematics, Universidad Carlos III de Madrid, Avda.\ Universidad 30; E-28911 Legan\'es, Spain\\
Tel.: +34-91-6249445\\
              Fax: +34-91-6249129\\
              \email{bonilla@ing.uc3m.es}           
\and
A. Carpio \at Departmento de Matem\'atica Aplicada, Universidad Complutense de Madrid; E-28040 Madrid, Spain}
\date{Received: \today / Accepted: date} 

\maketitle 

\begin{abstract}
We use our theory of periodized discrete elasticity to characterize  
defects in graphene as the cores of dislocations or groups of dislocations. Earlier numerical implementations of the theory predicted some of the simpler defect groupings observed in subsequent Transmission Electron Microscope experiments. Here we derive the more complicated defect groupings of three or four defect pairs from our theory, show that they correspond to the cores of two pairs of dislocation dipoles and ascertain their stability.
\keywords{Graphene \and Dislocations \and Periodized Discrete Elasticity}
\PACS{61.72.Bb \and 05.40.-a \and 61.48.De}
\end{abstract}

\section{Introduction}
\label{sec:1}
Graphene is a two-dimensional (2D) allotrope of carbon formed by a single layer of graphite that was first synthesized in 2004 \cite{nov04}. Since then, graphene has attracted great attention due to the Dirac-like spectrum of its charge carriers and the resulting extraordinary electronic properties \cite{gei07,cas09}. This fascinating material has remarkable electronic and mechanical properties \cite{gei07,cas09,voz10}. The electronic, chemical, thermal and mechanical properties of graphene are exceptionally sensitive to lattice imperfections \cite{gei07,cas09} and these defects and even the ripples that always cover suspended graphene sheets \cite{mey07,fas07} induce pseudo-magnetic gauge fields \cite{voz10}. Thus the study of defects in graphene is crucial and it has generated important experimental work \cite{mey08,wan08,col08,gom10}. Among observed defects, there are pentagon-heptagon (5-7) pairs, Stone-Wales (SW) defects (5-7-7-5 defects) \cite{mey08}, pentagon-octagon-pentagon (5-8-5) divacancies \cite{col08}, asymmetric vacancies (nonagon-pentagon or 9-5 pairs) and more complicated groupings such as 5-7-7-5 and 7-5-5-7 adjacent pairs or defects comprising three pentagons, three heptagons and one hexagon \cite{mey08}. In other two dimensional (2D) crystals such as Boron Nitride (hBN) symmetric vacancies have been observed \cite{mey09}. 

Real time observation of defect dynamics is possible using Transmission Electron Microscopes (TEM) corrected for aberration that have single atom resolution \cite{mey08}. Defect dynamics in graphene occurs on a time scale of seconds \cite{mey08}, much longer than sub-picosecond time scales typical of sound propagation in a primitive cell. On this long time scale and for unstressed graphene, SW defects are unstable: their two 5-7 pairs glide towards each other and annihilate, and the same occurs to defects comprising three pentagons, three heptagons and one hexagon, whereas 5-7-7-5 and 7-5-5-7 adjacent pairs remain stable \cite{mey08}. In stressed graphene oxide samples, SW defects split into their component 5-7 pairs which then move apart \cite{gom10}. While most theoretical studies on the influence of defects in electronic properties assume a given defect configuration and then proceed to analyze its effects \cite{cas09,voz10}, it is important to predict defect stability and evolution. 

In recent work, we have explained the observed long time defect dynamics in graphene by considering defects as the core of edge dislocations or dislocation dipoles in a planar 2D hexagonal lattice \cite{CBJV08,car08}. Our theory is a top-down approach whose starting point is linear elasticity. We discretize continuum linear elasticity on a hexagonal lattice and replace differences of vector displacements along primitive directions by periodic functions thereof which are linear for small differences. Our {\em periodized discrete elasticity} allows dislocation gliding along primitive directions and it reduces to continuum linear elasticity very far from dislocation cores \cite{car05}. Introducing a large damping in the resulting equations of motion and solving them numerically, we are able to predict the stable cores corresponding to a given dislocation configuration. Using this theory, we have predicted the stability of 5-7 defects (that are the cores of dislocations) \cite{CBJV08,car08}. Similarly, a study of dislocation dipoles in unstressed samples \cite{CBJV08,car08} predicts that SW are unstable whereas symmetric vacancies, divacancies and 7-5-5-7 defects are stable. In stressed samples, our theory predicts that SW split into two 5-7 pairs that move apart \cite{car08}, as confirmed later by experiments \cite{gom10}. 

In this paper, our theory is used to explain the evolution of defects involving a pair of 5-7-7-5 and 7-5-5-7 defects and a defect comprising three heptagons, three pentagons and one hexagon as observed by Meyer et al \cite{mey08}. 

The rest of the paper is as follows. Our theory and its equations of motion are explained in Section \ref{sec:2}. The stable cores corresponding to the far field of a single edge dislocation and a single dislocation dipole are used in Section \ref{sec:3} to illustrate the way defects are constructed numerically. Our results are also compared to available experiments in graphene and other 2D crystals. 
Section \ref{sec:4} contains the new results on more complex defects comprising two dislocation dipoles. We explain experimental observations by Meyer et al \cite{mey08}. The last section is devoted to our conclusions. 

\setcounter{equation}{0}
\section{Periodized discrete elasticity of planar graphene}
\label{sec:2}

\begin{figure}
\centering
\includegraphics[width=6cm]{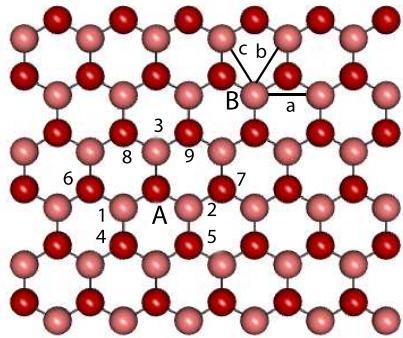}
\caption{(Color online) Structure of a graphene lattice. Neighbors of a given atom $A$ in sublattice 1 (dark red). Atoms in sublattice 2 are pink. The primitive vectors ${\bf a}$, ${\bf b}$ and ${\bf c}$ are also indicated. }
\label{fig1}
\end{figure}

In this paper we consider a planar graphene sample and ignore possible vertical deflections. In the continuum limit, in-plane deformations are described by the Navier equations of linear elasticity for the two-dimensional (2D) displacement vector $(u,v)$  \cite{ll7}. Including a phenomenological damping with coefficient $\gamma$ (to be fitted to experiments), we have 
\begin{eqnarray}
 \rho_2 \frac{\partial^2 u}{\partial t^2}+\gamma\frac{\partial u}{\partial t}= (\lambda+2\mu)\, \frac{\partial^2 u}{\partial x^2} +\mu\,\frac{\partial^2 u}{\partial y^2} + (\lambda+\mu)\,\frac{\partial^2 v}{\partial x \partial y}, \label{e1}\\
 \rho_2\frac{\partial^2 v}{\partial t^2}+\gamma\frac{\partial v}{\partial t}= \mu\,\frac{\partial^2 v}{\partial x^2} +
(\lambda+2\mu)\,\frac{\partial^2 v}{\partial y^2} +
 (\lambda +\mu)\,\frac{\partial^2 u}{\partial x \partial y}, \label{e2}
 \end{eqnarray}
where $\rho_2$ is the 2D mass density and $\lambda$ and $\mu$ are the 2D Lam\'e coefficients.  
 
 The governing equations of our theory are obtained in a three step process \cite{car08}: (i) discretize (1) on the hexagonal graphene lattice, (ii) rewrite the discretized equations in primitive coordinates, and (iii) replace finite differences appearing in the equations by periodic functions thereof in such a way that the equations remain invariant if we displace the atoms one step along any of the primitive directions. The last step allows dislocation gliding. 

 \subsection{Step (i): Discrete elasticity}
Next we discretize the equations of motion on a hexagonal lattice using the same notation as in Ref. \cite{car08}. Let us assign the coordinates $(x,y)$ to the atom $A$ in sublattice 1 (see Figure \ref{fig1}). The origin of coordinates, $(0,0)$, is also an atom of sublattice 1 at the center of the graphene sheet. The three nearest neighbors of $A$ belong to sublattice 2 and their cartesian coordinates are $n_1$, $n_2$ and $n_3$ below. Its six next-nearest neighbors belong to sublattice 1 and their cartesian coordinates are $n_i$, $i=4,\ldots, 9$:
\begin{eqnarray}
&& n_{1}=\left(x-{a\over 2},y-{a\over 2\sqrt{3}}\right), \, n_{2}=\left(x+{a\over 2},
y-{a\over 2\sqrt{3}}\right), \, n_{3}=\left(x,y+{a\over\sqrt{3}}\right),\nonumber\\
&& n_{4}=\left(x-{a\over 2},y-{a\sqrt{3}\over 2}\right), \, n_{5}=
\left(x+{a\over 2},y-{a\sqrt{3}\over 2}\right), \, n_{6}=(x-a,y), 
\nonumber\\
&& n_{7}=(x+a,y),\, n_{8}=\left(x-{a\over 2},y+{a\sqrt{3}\over 2}\right),\, 
n_{9}=\left(x+{a\over 2},y+{a\sqrt{3}\over 2}\right).\label{e3}
\end{eqnarray}
In Fig. \ref{fig1}, atoms $n_6$ and $n_7$ are separated from $A$ by the primitive vector $\pm {\bf a}$ and atoms $n_4$ and $n_9$ are separated from $A$ by the primitive vector $\pm {\bf b}$. Instead of choosing the primitive vector $\pm {\bf b}$, we could have selected the primitive direction $\pm {\bf c}$ along which atoms $n_8$, $A$ and $n_5$ lie. Let us define the following operators acting on functions of the coordinates $(x,y)$ of node $A$:
\begin{eqnarray}
Tu &=& [u(n_{1})-u(A)] + [u(n_{2})-u(A)] + [u(n_{3})-u(A)]\sim \left(\partial_x^2 u + \partial_y^2 u\right) {a^2\over 4},\label{T}\\
Hu & = & [u(n_{6})-u(A)]+[u(n_{7})-u(A)]\sim (\partial_x^2 u)\,  a^2,  \label{H} \\
D_{1}u &=& [u(n_{4})-u(A)] + [u(n_{9})-u(A)]\sim \left({1\over 4}\, \partial_x^2 u +{\sqrt{3}\over 2}\,\partial_x\partial_yu + {3\over 4}\, \partial_y^2 u\right)a^2,  \label{D1}\\
D_{2}u &=& [u(n_{5})-u(A)] + [u(n_{8})-u(A)]\sim \left({1\over 4}\, \partial_x^2 u -{\sqrt{3}\over 2}\,\partial_x\partial_yu + {3\over 4}\, \partial_y^2 u\right)a^2,  \label{D2}
 \end{eqnarray}
as the lattice constant $a$ tends to zero. Similar operators can be defined if we replace the point $A$ in sublattice 1 by a point belonging to the sublattice 2. Now we replace in (\ref{e1}) and (\ref{e2}), $Hu/a^2$, $(4T-H)u/a^2$ and $(D_{1}-D_{2})u/(\sqrt{3}a^2)$ instead of $\partial_x^2 u$, $\partial_y^2 u$ and $\partial_x\partial_yu$, respectively, with similar substitutions for the derivatives of $v$, thereby obtaining the following equations at each point of the lattice:
\begin{eqnarray}
\rho_2 a^2 \partial_t^2 u + \gamma\,\partial_t u&=& 4\mu\, Tu + (\lambda+\mu)\, Hu 
+{\lambda+\mu \over\sqrt{3}}\, (D_{1} - D_{2})v, \label{e4}\\
\rho_2 a^2 \partial_t^2 v + \gamma\,\partial_t v&=& 4 (\lambda+2\mu )\, Tv - (\lambda+\mu )H v +{\lambda+\mu\over\sqrt{3}}\, (D_{1} - D_{2})u . \label{e5}
 \end{eqnarray}
 These equations have two characteristics time scales, the time $t_s=\sqrt{\rho_2 a^2/(\lambda+2\mu)}$ it takes a longitudinal sound wave to traverse a distance $a$ and the characteristic damping time, $t_d=\gamma a^2/(\lambda+2\mu)$. Using the known values of the Lam\'e coefficients at the graphite basal plane \cite{bla70} \footnote{At 300 K, $C_{66}=\mu_{3D}= 440$ GPa, $C_{12}= \lambda_{3D}= 180$ GPa, $C_{11}=\lambda_{3D}+2\mu_{3D}=1060$ GPa. The 2D coefficients are $\lambda=\lambda_{3D}d$ and $\mu=\mu_{3D}d$, where $d=$ 3.35 \AA\, is the distance between graphene planes in graphite. Similarly, $\rho_2=\rho d$ is found from the bulk density of graphite.} (that agree with calculations \cite{zak09} and measurements in graphene \cite{lee08}), $t_s\approx 10^{-14}$ s. Our simulations show that it takes $0.4 t_d$ a SW to disappear after it is created by irradiation which, compared with the measured time of 4 s \cite{mey08}, gives $t_d\approx 10$ s. On a $t_d$ time scale, we can ignore inertia in (\ref{e4})-(\ref{e5}).

\subsection{Step (ii): Nondimensional equations in primitive coordinates}
We now transform (\ref{e4})-(\ref{e5}) to the nondimensional primitive coordinates $u'$, $v'$ using $u=a(u'+v'/2)$, $v=\sqrt{3}av'/2$, use the nondimensional time scale $t'=t/t_d$ and ignore inertia. The resulting equations are
\begin{eqnarray}
&&\partial_{t'} u'= \frac{4\mu Tu'}{\lambda+2\mu} +
\frac{\lambda +\mu}{\lambda+2\mu}\left[\left(H-\frac{D_{1}-D_{2}}{3}\right)u' + \left(H+\frac{D_{1}-D_{2}}{3}-2T\right)v'\right], \label{e10} \\
&&\partial_{t'} v' = \frac{2}{3}\,\frac{\lambda +\mu}{\lambda+2\mu}\, (D_{1}-D_{2})u'+ 4Tv' + \frac{\lambda +\mu}{\lambda+2\mu}\left(\frac{D_1-D_{2}}{3}-H\right)v' . \label{e11}
\end{eqnarray}

\subsection{Step (iii): Periodized discrete elasticity}
The models described by the linear equations (\ref{e10}) - (\ref{e11}) do not allow for the changes of neighbors involved in defect motion. One way to achieve these changes is to update neighbors as a defect moves. Then (\ref{e10}) and (\ref{e11}) would have the same appearance, but the neighbors $n_i$ would be given by (\ref{e3}) only at the start. At each time step, we keep track of the position of the different atoms and update the coordinates of the $n_{i}$. This is commonly done in Molecular Dynamics, as computations are actually carried out with only a certain number of neighbors. Convenient as updating is, its computational cost is high and analytical studies thereof are not easy. 

In simple geometries, we can avoid updating by introducing a periodic function of differences in the primitive directions that automatically describes link breakup and union associated with defect motion. Besides greatly reducing computational cost, the resulting periodized discrete elasticity models allow analytical studies of defect depinning \cite{car03,car05}, motion and nucleation \cite{pcb08,pcb09}. Another advantage of periodized discrete elasticity is that boundary conditions can be controlled efficiently to avoid spurious numerical reflections at boundaries.

To restore crystal periodicity, we replace the linear operators $T$, $H$, $D_{1}$ and $D_{2}$ in (\ref{e10}) and (\ref{e11}) by their periodic versions:
\begin{eqnarray}
&& T_{p}u'=g(u'(n_{1})-u'(A)) + g(u'(n_{2})-u'(A)) + g(u'(n_{3})-u'(A)),\nonumber\\
&& H_{p}u'= g(u'(n_{6})-u'(A))+g(u'(n_{7})-u'(A)), \nonumber\\
&& D_{1p}u'=g(u'(n_{4})-u'(A)) + g(u'(n_{9})-u'(A)),\nonumber\\
&& D_{2p}u'=g(u'(n_{5})-u'(A)) + g(u'(n_{8})-u'(A)),\label{e12}
\end{eqnarray}
where $g$ is a periodic function, with period one, and such that $g(x)\sim x$ as $x\to 0$. We obtain:
\begin{eqnarray}
&&\partial_{t'} u'= \frac{4\mu T_pu'}{\lambda+2\mu} +
\frac{\lambda +\mu}{\lambda+2\mu}\left[\left(H_p-\frac{D_{1p}-D_{2p}}{3}\right)u' + \left(H_p+\frac{D_{1p}-D_{2p}}{3}-2T_p\right)v'\right], \label{e6} \\
&&\partial_{t'} v' = \frac{2}{3}\,\frac{\lambda +\mu}{\lambda+2\mu}\, (D_{1p}-D_{2p})u'+ 4T_pv' + \frac{\lambda +\mu}{\lambda+2\mu}\left(\frac{D_{1p}-D_{2p}}{3}-H_p\right)v' . \label{e7}
\end{eqnarray}

In our tests we have taken $g$ to be a periodic piecewise linear continuous function: 
\begin{eqnarray}\label{e13}
g_{\alpha}(x)= \left\{ 
\begin{array}{ll}
x, & -\alpha \leq x \leq \alpha, \\
-{2 \alpha \over 1 - 2 \alpha}  x + {\alpha \over 1 - 2 \alpha},
  &  \alpha \leq x \leq 1 -\alpha.
\end{array}\right.
\end{eqnarray}
The parameter $\alpha$ controls defect stability and mobility under applied stress. It 
should be sufficiently large for elementary defects (dislocations, vacancies) to be stable at zero applied stress, and sufficiently small for dislocations to move under reasonable applied stress \cite{car05}. We use $\alpha=0.4$ to account for experimentally observed
stability properties of the defects. For lower values, the stable defect described in section \ref{sec:4} loses the Stone-Wales component. The periodic function $g$ can be replaced by a different type of periodic function to achieve a better fit to available experimental or numerical data.

\section{Stable cores of dislocations and dislocation dipoles}
\label{sec:3}
\subsection{Boundary and initial conditions for a single dislocation}
We solve (\ref{e6})-(\ref{e7}), or (\ref{e10})-(\ref{e11}) with the periodic operators $T_p$, $H_p$, $D_{1p}$ and $D_{2p}$, using as initial and boundary conditions the far field of appropriate dislocations which are the stationary solutions of the linear elasticity equations \cite{ll7}. Since the latter are a good approximation four spacings away from the core of SW defects in graphene, and our model equations seamlessly reduce to linear elasticity in the far field, we use a relatively small lattice with $18\times 18$ spacings ($36\times 36$ carbon atoms) in our numerical simulations \cite{car08}. Consider first the case of a single edge dislocation with Burgers vector $(a,0)$ and displacement vector ${\bf u}= (u(x,y),v(x,y))$

\begin{figure}
\centering
\includegraphics[width=4cm]{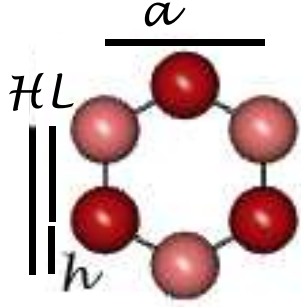}
\caption{(Color online) To generate defects, we use lengths depicted in the figure as referred to a hexagon cell: $a$ and $l=a/\sqrt{3}$ are the lattice constant and the hexagon side, respectively. $H=l+h$, where $h=l/2$ is the vertical distance between nearest neighbor atoms.}
\label{fig1bis}
\end{figure}

\begin{eqnarray} 
u &= & \frac{a}{2\pi} \left[\tan^{-1}\left({y\over x}\right) + 
{ xy \over 2(1-\nu)(x^2+y^2)}\right],\nonumber\\  
v &= & \frac{a}{2\pi} \left[ -{1-2\nu\over 4(1-\nu)}\,\ln\left( 
{x^2+y^2\over a^2}\right) + {y^2 \over 2(1-\nu) (x^2+y^2)}\right], \label{e14}
\end{eqnarray}
where $\nu=\lambda/[2(\lambda+\mu)]$ is dimensionless; cf.\ Ref.~\cite{ll7}, pag.\ 114. (\ref{e14}) has a singularity $\propto (x^2+y^2)^{-1/2}$ at the origin of coordinates and it satisfies $\int_{\mathcal{ C}} (d{\bf x}\cdot\nabla) {\bf u} =- (a,0)$, for any closed curve $\mathcal{C}$ encircling the origin. Using (\ref{e14}), we write $\mathbf{u}=(u,v)$ in primitive coordinates, $U'(l,m)= [u(x-x_0,y-y_0)-v(x-x_0,y-y_0)/\sqrt{3}]/a$, $V'(l,m)=2v(x-x_0,y-y_0)/(a\sqrt{3})$, where $x=(x'+y'/2)a$, $y=\sqrt{3}\, ay'/2$, $x'=l$, $y'=m$ (integers) and $(x_0,y_0)\neq (0,0)$ to avoid that the singularity in (\ref{e14}) be placed at a lattice point. To find defects, we solve the periodized discrete elasticity equations (\ref{e6})-(\ref{e7}) with the initial and boundary conditions:
\begin{eqnarray}
\mathbf{u'}(l,m;0)=\mathbf{U'}(l,m),\quad\mbox{and}\quad 
\mathbf{u'}(l,m;t)=\mathbf{U'}(l,m)+ F(m,0) \quad\mbox{at lattice boundaries.} \label{e8}
\end{eqnarray}
Here $F$ is a dimensionless applied shear stress. For $|F|<F_c$ (Peierls stress), the solution of (\ref{e6})-(\ref{e7}) relaxes to a stable dislocation $(u'(l,m),v'(l,m))$ with appropriate far field, which is (\ref{e14}) if $F=0$. 

Numerical simulations give us the location of carbon atoms at each time $t$. We represent atoms by spheres of arbitrary size. As a guide to the eye and to visualize defects more easily, we have attached fictitious bonds to these spheres \cite{CBJV08,car08}. Depending on the location of the singularity $(x_0,y_0)$, there are two possible configurations corresponding to the same edge dislocation in the continuum limit. If the singularity is placed between two atoms that form any non-vertical side of a given hexagon, the core of the deformed lattice $(l+u'(l,m),m+v'(l,m))$ is a 5-7 (pentagon-heptagon) defect. If the singularity is placed in any other location different from a lattice point, the core of the singularity forms an octagon having one atom with a dangling bond \cite{CBJV08,car08}. Stable 5-7 defects are commonly observed in experiments \cite{mey08,mey09,gom10}, whereas adsorbed atoms (not considered in our model) may attach to a dangling bond thereby destroying the octagon configuration. 
\begin{figure}
\centering
\includegraphics[width=8cm]{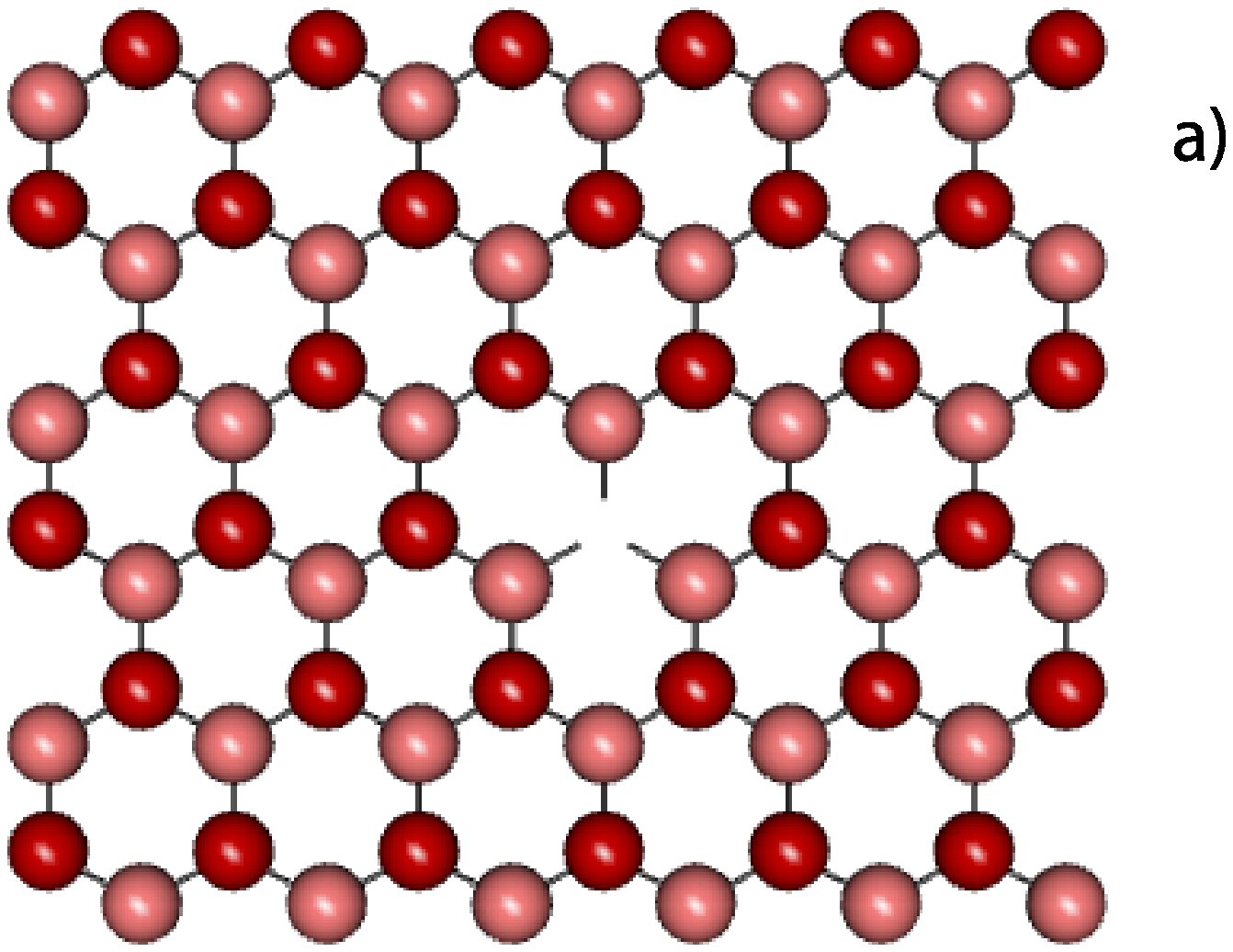}
\includegraphics[width=8cm]{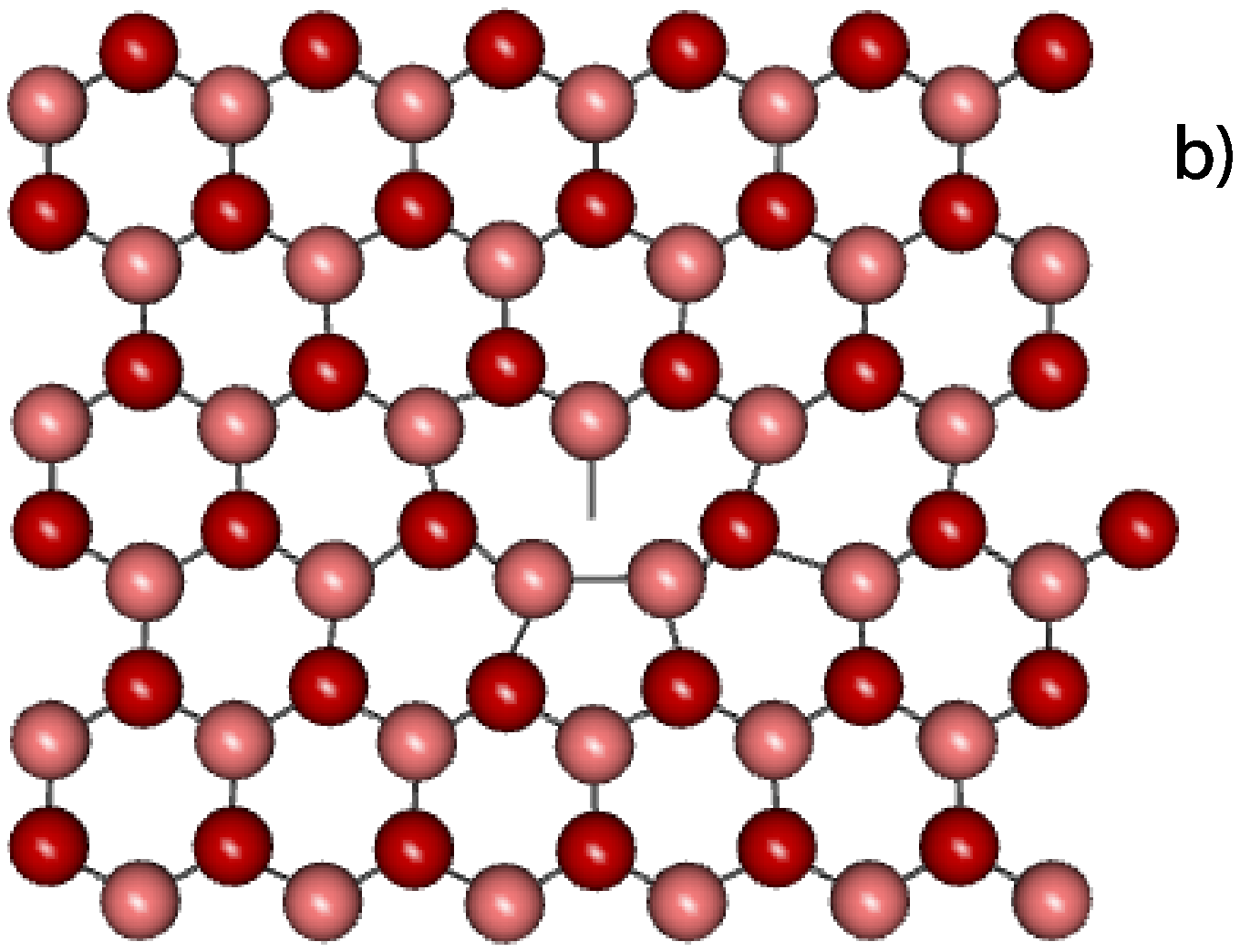}
\caption{(Color online) (a) Symmetric vacancy. (b) Asymmetric vacancy (nonagon-pentagon defect).}
\label{fig2}
\end{figure}

\begin{figure}
\centering
\includegraphics[width=8cm]{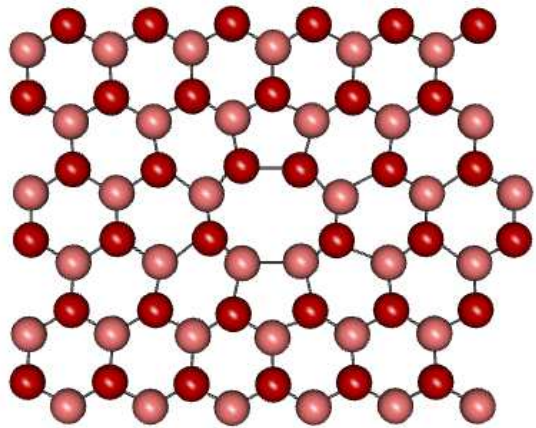}
\caption{(Color online) Pentagon-octagon-pentagon divacancy.}
\label{fig3}
\end{figure}

\begin{figure}
\centering
\includegraphics[width=8cm]{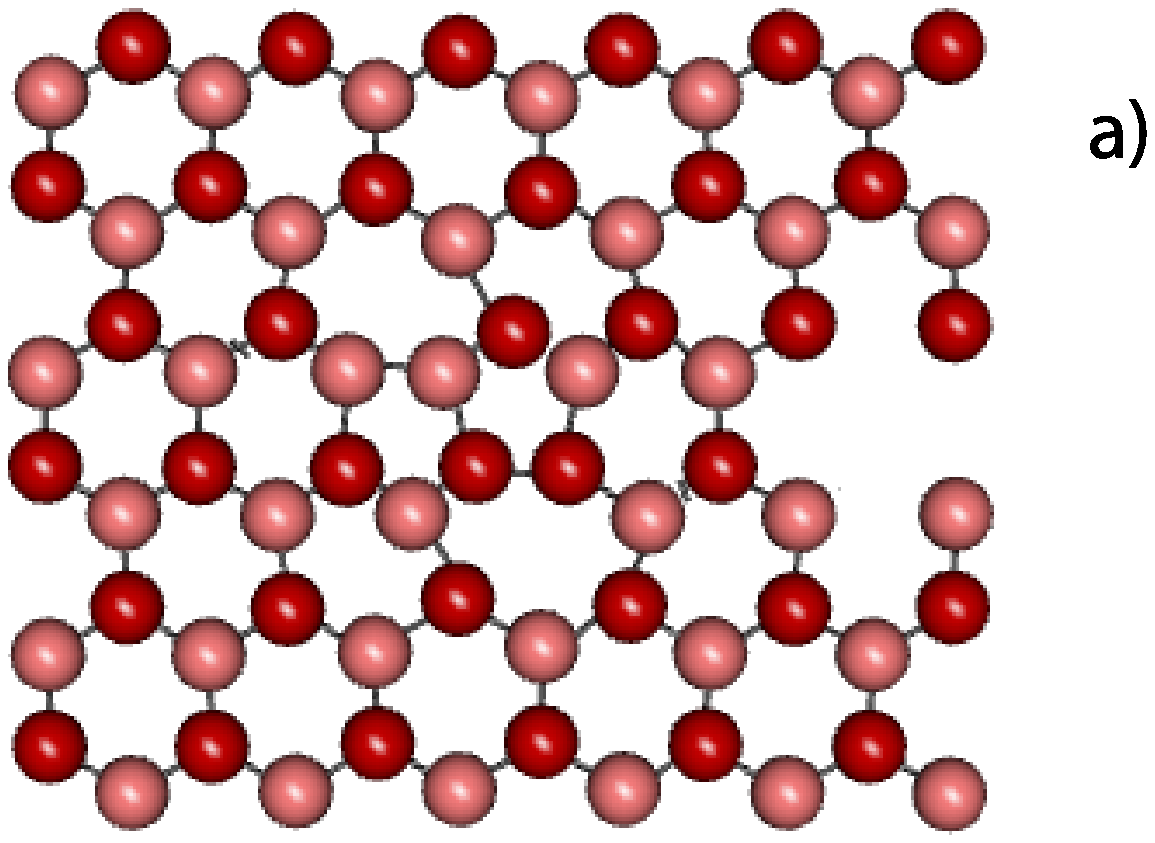}
\includegraphics[width=8cm]{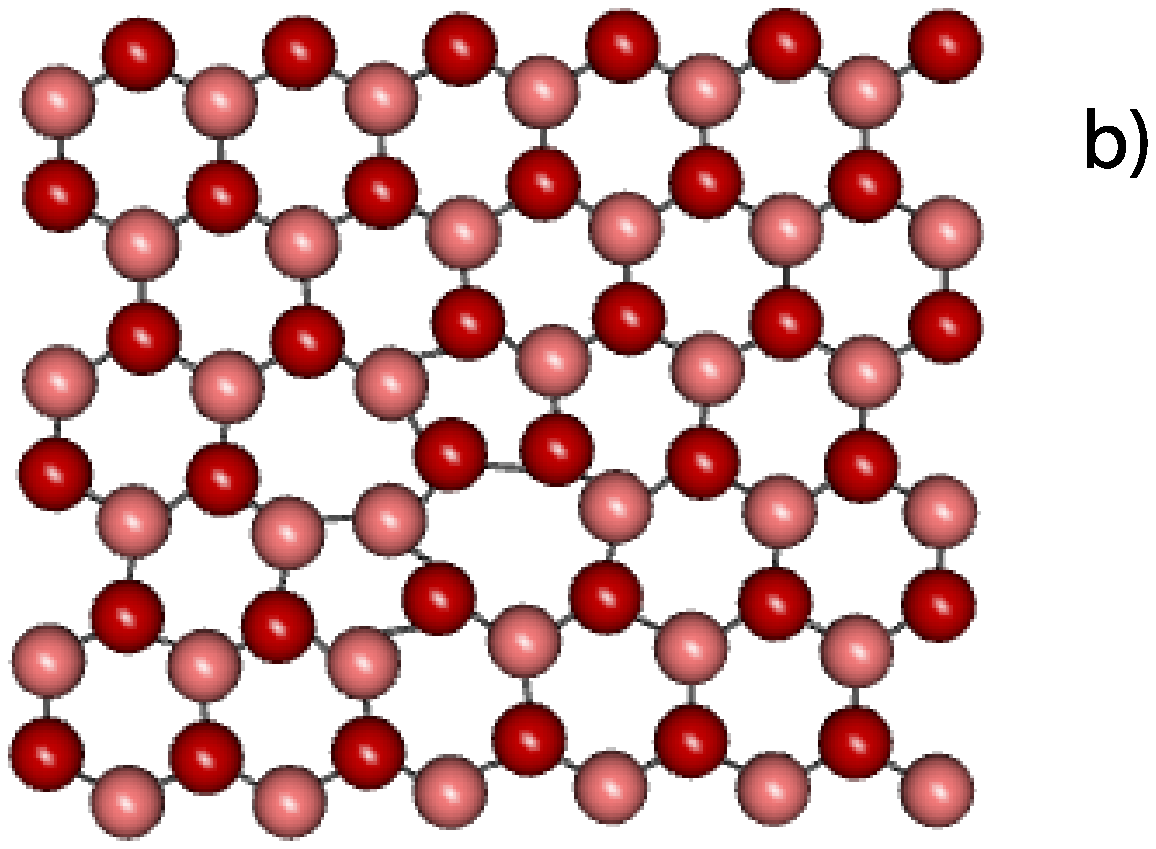}
\caption{(Color online) (a) Stable 7-5-5-7 defect. (b) Unstable 5-7-7-5 Stone Wales defect.}
\label{fig4}
\end{figure}

\subsection{Dislocation dipoles}
A dislocation dipole is formed by two dislocations with opposite Burgers vectors, $\pm\mathbf{a}$. Depending on how we place the origin of coordinates, different dipole configurations result. Let ${\bf U}(x,y)$ be the displacement vector (\ref{e14}) of a single dislocation. We find the dipole cores by selecting as zero stress initial and boundary conditions ${\bf U}(x-x_{0}^+,y-y_{0}^+)-{\bf U}(x-x_{0}^-,y-y_{0}^-)$, with different $(x^\pm_0,y_0^\pm)$. Let $a$, $l=a/\sqrt{3}$, $H=3l/2$ and $h=l/2$ be the lattice constant, the hexagon side, the vertical distance between two nearest neighbor atoms belonging to the same sublattice, and the vertical distance between nearest neighbor atoms having different ordinate, respectively; cf. Fig. \ref{fig1bis}. We get:
\begin{itemize}
\item Vacancies: $x_0^+=-0.25 a$, $y_0^+=-0.8 h+H/2$ and $x_0^-=-0.25 a$, $y_0^-=-0.8 h$. This initial configuration is the asymmetric vacancy (9-5 defect) of Fig. \ref{fig2}(b), which evolves to the symmetric vacancy of Fig. \ref{fig2}(a) under overdamped dynamics.
\item Stable divacancy: $x_0^+=-0.25 a$, $y_0^+=-0.8 h+ H$ and $x_0^-=-0.25 a$, $y_0^-=-0.8 h$. Fig. \ref{fig3}.
\item Stable 7-5-5-7 defect: $x_0^+=-0.25 a+a$, $y_0^+=-0.8 h$ and $x_0^-=-0.25 a$, $y_0^-=-0.8 h+H$. Fig. \ref{fig4}(a).  
\item Unstable Stone-Wales 5-7-7-5 defect: $x_0^+=-0.25 a+a$, $y_0^+=-0.8 h$ and $x_0^-=-0.25 a$, $y_0^-=-0.8 h$. Fig. \ref{fig4}(b). For $F=0$, this initial configuration corresponds to two dislocations with opposite Burgers vectors that share the same glide line, and it evolves to the undisturbed lattice when the dislocations move towards each other and annihilate.
\end{itemize}

\subsection{Comparison with results of experiments}
Carbon atoms and defects in graphene sheets are visualized by operating at low voltage ($\leq$ 80 kV, to avoid irradiation damage to the sample\footnote{For an 80
keV incident electron, the maximum energy that can be transferred to a carbon atom is 15.8 eV. This is below the threshold for knock-on damage (17 eV, corresponding to a beam energy of 86 keV) but sufficient to form multiple SW defects \cite{mey08,gir09}. Migration of carbon atoms to empty neighboring sites has a significantly lower cost and therefore dislocation motion is much easier. The actual temperature rise in the suspended graphene specimen due to exposure by the electron beam in the microscope is minimal because the beam
current density is small ($\sim$45 A/cm$^2$) and the thermal conductivity of graphene is extremely high ($>$1000 W/mK). Thus the sample is not far from room temperature during the experiments \cite{gir09}.}) a transmission electron aberration-corrected microscope (TEAM) with appropriate optics \cite{mey08}. This microscope is capable of sub-\AA ngstrom resolution even at 80 kV and can produce real time images of carbon atoms on a scale of seconds: each frame averages 1s of exposure and the frames themselves are 4 s apart \cite{mey08,gir09}. The images obtained in experiments can be used to determine the time evolution of defects in graphene created by irradiation or sample treatment \cite{mey08,mey09,gom10}. 

In experiments, both symmetric and asymmetric vacancies are observed in unstressed graphene \cite{mey08}, whereas in single layers of hexagonal Boron Nitride (hBN) only symmetric vacancies are observed \cite{mey09}. Stable 5-8-5 divacancies are also observed \cite{col08}. The annihilation of the 5-7-7-5 SW defect in \ref{fig4}(b) (the heptagons share one side) 4 s after its creation is seen in Figures 3(c) and (d) of \cite{mey08}. Our model predicts that SW under sufficient strain split in their two 5-7 pairs that move apart (cf Fig 6 of \cite{car08}), which has been observed very recently; cf Fig. 4(a) and (b) in \cite{gom10}.
\begin{figure}
\centering
\includegraphics[width=8cm]{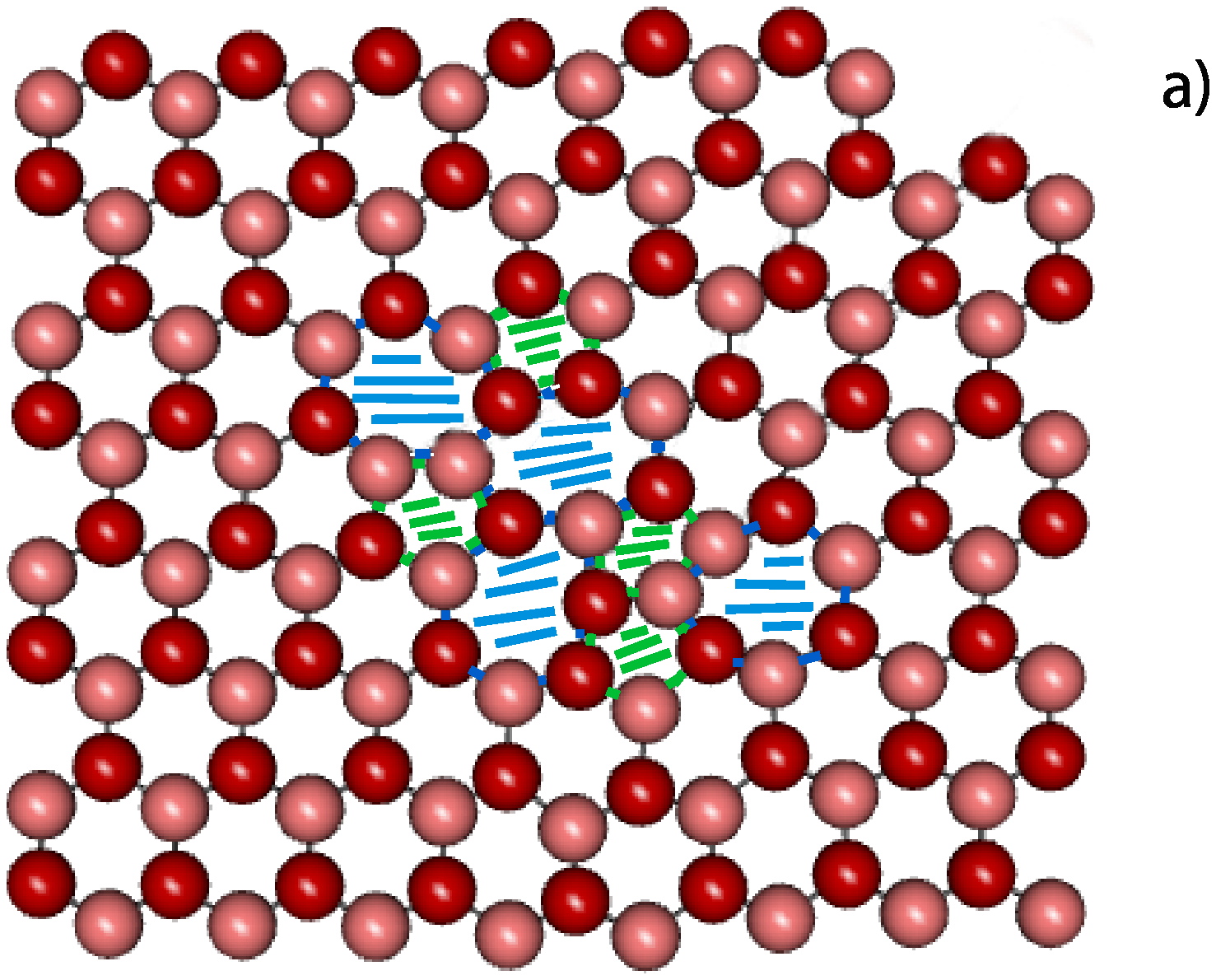}
\includegraphics[width=8cm]{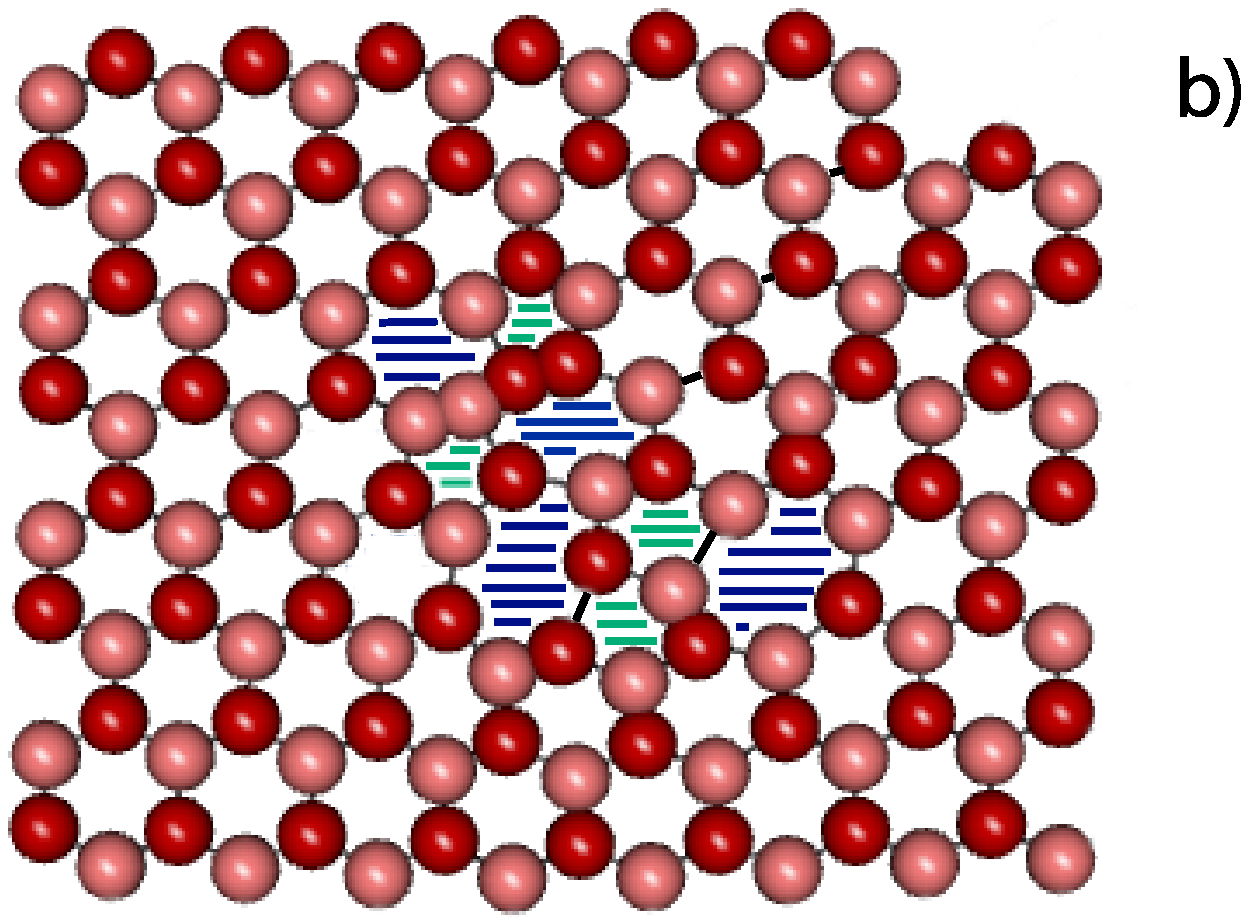}
\caption{(Color online) Defects corresponding to two dislocation dipoles: (a) Initial configuration of a pair of 5-7-7-5 and 7-5-5-7 defects. (b) Final configuration. } 
\label{fig5}
\end{figure}

\section{Dislocation dipole pairs}
\label{sec:4}
We study the evolution of configurations comprising two dislocation dipoles each in order to explain experimental observations by Meyer et al \cite{mey08}. \ref{fig5}(a) depicts an initial condition consisting of a 5-7-7-5 SW defect adjacent to a rotated 7-5-5-7 defect ($a$, $l=a/\sqrt{3}$, $h=l/2$ and $H=3l/2$ are the distances indicated in Figure \ref{fig1bis}):
\begin{eqnarray} 
&&{\bf U}(x-x_{0}^+,y-y_{0}^+)-{\bf U}(x-x_{0}^-,y-y_{0}^-)
+{\bf \tilde{U}}(\tilde{x}-\tilde{x}^+_{0},\tilde{y}-\tilde{y}_{0}^+)-{\bf \tilde{U}}(\tilde{x}-\tilde{x}_{0}^-,\tilde{y} -\tilde{y}_{0}^-),\label{4.1}\\
&&x_{0}^+=-0.3 a,\, y_{0}^+= -0.7 h+2H,\quad x_0^-=-0.3a- a,\, y_0^-=-0.7h+2H,\nonumber\\ 
&&\tilde{x}_0^+=0.3 a+a,\,\tilde{y}_0^+=0.3h-H,\quad\tilde{x}_0^-=0.15a,\, \tilde{y}_0^+ =0.5 h.\label{4.2}
\end{eqnarray} 
Here ${\bf U}(x,y)$ is the edge dislocation (\ref{e14}) with origin of coordinates at a central atom of type $A$ in Figure \ref{fig1} and Burgers vector $\mathbf{a}$ (in units of the lattice constant $a$). ${\bf \tilde{U}}(\tilde{x},\tilde{y})$ is an edge dislocation with Burgers vector $\mathbf{b}$. To obtain ${\bf \tilde{U}}(\tilde{x},\tilde{y})$, we first consider the axes $(\tilde{x},\tilde{y})$ rotated a $\pi/3$ angle from the axes $(x,y)$. Next we form a 7-5-5-7 defect by combining a {\em positive} dislocation with Burgers vector $(a,0)$ centered at $(\tilde{x}_0^+,\tilde{y}^+_0)$ and a {\em negative} dislocation with Burgers vector $(-a,0)$ centered at $(\tilde{x}_0^-,\tilde{y}^-_0)$. Then the result is rewritten in the original coordinates $(x,y)$. Now we add the 5-7-7-5 SW defect given by the first two terms in (\ref{4.1}) and complete that equation. The initial and boundary condition (\ref{4.1}) of the defect correspond to two dislocation dipoles having Burgers vectors along two different primitive directions and it is the same one as reported in Figures 3(h) and (i) of Meyer et al's experiments \cite{mey08}. Under overdamped dynamics, this defect remains stable. As predicted in \cite{CBJV08,car08}, the 7-5-5-7 defect is stable and this apparently stabilizes our pair of dislocation dipoles for the selected initial configuration. Other nearby configurations evolve to two octagons corresponding to a dipole comprising two edge dislocations with opposite Burgers vectors. As explained before, adsorbed atoms may be attached to the dangling bonds thereby eliminating these configurations and restoring the undisturbed hexagonal lattice.

\begin{figure}
\centering
\includegraphics[width=8cm]{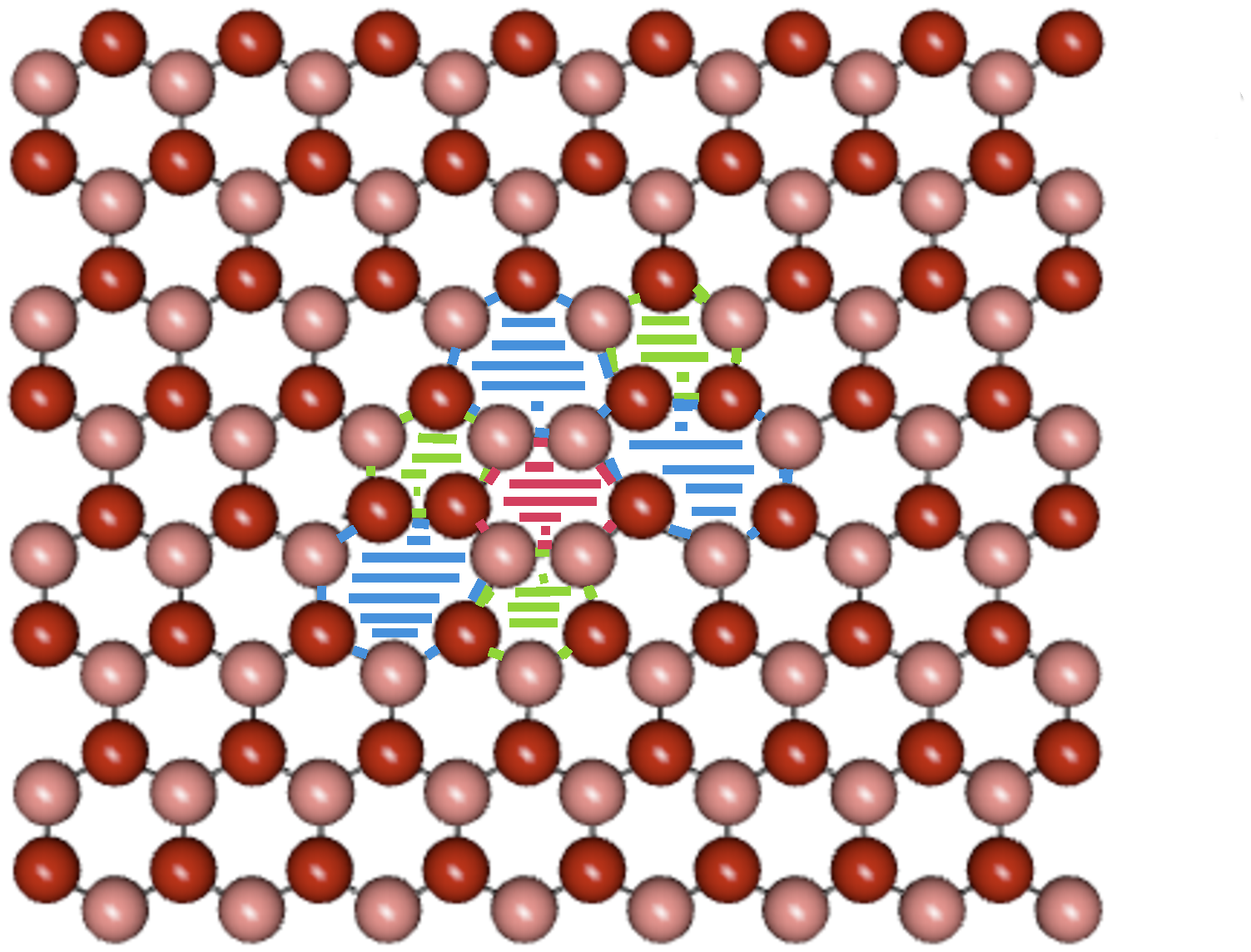}
\caption{(Color online) Metastable defect comprising three heptagons, three pentagons and one hexagon. It evolves to the defect-free lattice after 4 s.}
\label{fig6}
\end{figure}

The other hitherto unexplained defect configuration in Figures 3(j) and (k) of Ref. \cite{mey08} is a metastable defect consisting of three pentagons and three heptagons. This defect appears with the following initial and boundary condition depicted in Figure \ref{fig6}: 
\begin{eqnarray} 
&&{\bf U}(x-x_{0}^+,y-y_{0}^+)-{\bf U}(x-x_{0}^-,y-y_{0}^-) 
+{\bf U}(x-\hat{x}^+_{0},y-\hat{y}_{0}^+)-{\bf U}(x-\hat{x}_{0}^-,y -\hat{y}_{0}^-),\label{4.3}\\
&&x_{0}^+=-0.3 a+a,\,\, y_{0}^+= -0.7 h,\quad x_0^-=-0.3 a,\,\, y_0^-=-0.7 h,\nonumber\\ 
&&\hat{x}_0^+=-0.3 a-a,\,\,\hat{y}_0^+=-0.7h-H,\quad\hat{x}_0^-=-0.3 a,\,\,\hat{y}_0^+= - 0.7h-H,\label{4.4}
\end{eqnarray}
corresponding again to two dislocation dipoles all whose component dislocations have Burgers vectors directed along the $x$ axis. Starting from a negative dislocation centered at $(x_0^-,y_0^-)=(-0.3a,-0.7h)$, the first dipole adds a positive dislocation shifted one lattice constant to the left. The second dipole consists of a negative dislocation shifted vertically upwards a distance $H=3l/2=\sqrt{3}a/2$ (1.5 times the hexagon side, or $\sqrt{3}/2$ times the lattice constant) from $(x_0^-,y_0^-)$ and a positive dislocation which shifts horizontally to the right the previous one a distance equal to one lattice constant. Under overdamped dynamics, this defect disappears as the positive and negative dislocations comprising each dipole glide towards each other. See the movie in the Supplementary material. This agrees with Meyer et al's experimental observation \cite{mey08}. 

\section{Conclusions}
\label{sec:5}
In summary, the proposed theory of defect dynamics in planar graphene regularizes continuum linear elasticity on a hexagonal lattice by replacing linear combinations of four appropriate difference operators acting on the displacement vector instead of partial derivatives thereof, and periodizes these operators along primitive directions. Far from defect cores where differences of the displacement vector are sufficiently small, the resulting discrete equations seamlessly reduce to those of continuum linear elasticity. Adding large damping terms, these equations are solved with appropriate initial and boundary conditions consistent with the known solutions corresponding to edge dislocations and edge dislocation dipoles in linear elasticity. The numerical solutions of these equations explain the stability and evolution of several experimentally observed defects in suspended graphene sheets. Observed isolated defects are the cores of edge dislocations, dislocation dipoles or pairs of dipoles. Among them, isolated dislocations (pentagon-heptagon pairs), dislocation dipoles (symmetric vacancies, nonagon-pentagon pairs which are asymmetric vacancies, 5-8-5 divacancies, 5-7-7-5 Stone Wales defects and 7-5-5-7 defects) and pairs of dislocation dipoles (a 5-7-7-5 SW defect adjacent to a 7-5-5-7 defect and a metastable defect comprising three pentagons, three heptagons and one hexagon). The theory correctly predicts stability or instability of these defects as observed in experiments. 

\acknowledgement
This work has been financed by the Spanish Ministry of Science and Innovation (MICINN) under grants FIS2008-04921-C02-01 (LLB), FIS2008-04921-C02-02 and UCM/BSCH CM 910143 (AC).

\end{document}